%% file: main.tex
\documentclass[conference]{IEEEtran}
\IEEEoverridecommandlockouts

\usepackage{cite}
\usepackage[tbtags]{amsmath}
\usepackage{amssymb,amsfonts}
\usepackage{graphicx}
\usepackage{stfloats}
\usepackage{float}
\usepackage{balance}
\usepackage{needspace}
\usepackage{textcomp}
\usepackage{xcolor}
\usepackage{url}
\usepackage{orcidlink}
\hypersetup{hidelinks}
\AtBeginDocument{%
  \setcounter{topnumber}{2}%
  \setcounter{bottomnumber}{2}%
  \setcounter{totalnumber}{4}%
  \setlength{\textfloatsep}{8pt plus 2pt minus 2pt}%
  \setlength{\dbltextfloatsep}{8pt plus 2pt minus 2pt}%
  \setlength{\floatsep}{8pt plus 2pt minus 2pt}%
  \setlength{\dblfloatsep}{8pt plus 2pt minus 2pt}%
  \setlength{\intextsep}{8pt plus 2pt minus 2pt}%
  \setlength{\abovedisplayskip}{4pt plus 2pt minus 1pt}%
  \setlength{\belowdisplayskip}{4pt plus 2pt minus 1pt}%
  \setlength{\abovedisplayshortskip}{0pt plus 2pt}%
  \setlength{\belowdisplayshortskip}{3pt plus 2pt minus 1pt}%
  \setlength{\jot}{2pt}%
}
\usepackage{tikz}
\usetikzlibrary{arrows.meta,positioning}

\title{\fontsize{22}{26}\selectfont
EdgeNet City: Context-Aware AI-RAN Planning\\
toward Autonomous Urban Connectivity
\thanks{This work was supported in part by the UK EPSRC grant EP/X038971/1 and the Horizon Europe grants 101086219 and 101212141. Research supported by the NVIDIA Academic Grant Program using NVIDIA-provided RTX PRO 6000 Blackwell Max-Q Workstation Edition GPUs. \newline This work has been submitted to the IEEE for possible publication. Copyright may be transferred without notice, after which this version may no longer be accessible.}
}

\author{
\IEEEauthorblockN{
Ziming Liu\,\orcidlink{0009-0002-6772-0005}\IEEEauthorrefmark{1},
Boxuan Shan\,\orcidlink{0009-0006-0329-9978}\IEEEauthorrefmark{1},
Qixing Deng\,\orcidlink{0009-0004-6815-3422}\IEEEauthorrefmark{2},
Wenchao Xue\,\orcidlink{0009-0004-4206-5790}\IEEEauthorrefmark{3},
and Xiaoli Chu\,\orcidlink{0000-0003-1863-6149}\IEEEauthorrefmark{1}
}
\IEEEauthorblockA{
\IEEEauthorrefmark{1}The University of Sheffield, Sheffield, UK \quad
\IEEEauthorrefmark{2}Imperial College London, UK \quad
\IEEEauthorrefmark{3}Independent Researcher, London, UK\\
\texttt{\{ziming.liu,bshan3,x.chu\}@sheffield.ac.uk}\\
\texttt{cindy.deng22@imperial.ac.uk};
\texttt{wincherx@126.com}
}
}

\begin{document}

\IEEEaftertitletext{\vspace{-12pt}}

\raggedbottom

\maketitle

\input{figures/fig_context_to_cow}

\input{sections/00_abstract}
\input{sections/01_introduction}
\input{sections/02_system_model_and_problem_formulation}
\input{sections/03_methodology}
\input{sections/04_results}

\input{sections/06_conclusion}
\input{sections/08_references}

\end{document}

%% file: figures/fig_context_to_cow.tex
\begin{figure*}[!t]
\centering
\includegraphics[width=\textwidth]{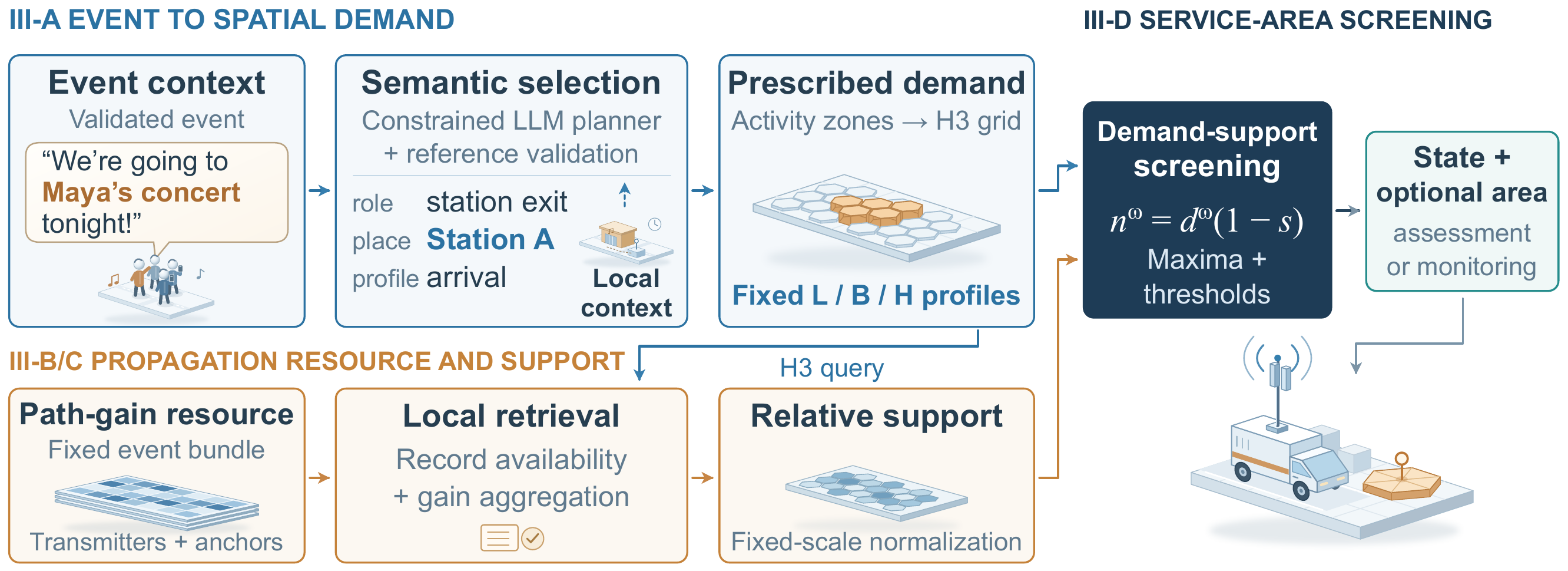}
\caption{Illustrative context-to-COW workflow. Validated activity selections prescribe H3 demand. Reused path-gain records determine cell eligibility and relative propagation support. Demand--support screening returns a state and an optional service area.}
\label{fig:context-to-cow}
\label{fig:method-dataflow}
\label{fig:context-to-cow-integrated}
\end{figure*}

%% file: sections/00_abstract.tex
\begin{abstract}
The evolution from 5G toward 6G calls for more anticipatory and
context-aware Cell-on-Wheels (COW) planning in wireless networks. We present EdgeNet City, an online framework that grounds free-form
natural-language inputs in geographic activity representations and
path-gain data. A catalog-referenced representation links activity
roles, geographic features and predefined demand profiles, separating
semantic selection from deterministic spatial computation. The algorithm
combines prescribed demand with relative propagation support to screen
service areas for subsequent COW assessment. We publicly release a London dataset
with 22,678 transmitters, approximately 449.4 billion
path-gain records and retained receiver geometry. Within prepared
propagation domains, the online prototype reuses these records without new
ray tracing or live operator measurements. In a matched 20-event study
across six London environments, the expanded representation increases mean
spatial overlap with the reused full technical-review reference from 0.4929
for the single-feature Venue-only baseline to 0.7203. Rule-based and
language-model planners select identical activities in this panel. With
need-aware ranking and a three-cell diagnostic budget fixed, the expanded
representation increases mean Reference-need Capture by 25.44 percentage
points over Venue-only.
\end{abstract}

\begin{IEEEkeywords}
Cell-on-Wheels, service-area screening, event-induced demand,
urban propagation datasets, activity representation
\end{IEEEkeywords}

%% file: sections/01_introduction.tex
\section{Introduction}
In 5G and future 6G wireless networks, context-aware radio planning must
locate anticipated activities described in natural language and assess their
demand against local propagation conditions. Urban events can
concentrate cellular demand, changing traffic volumes, active-user counts
and connection failures \cite{shafiq2013}. Schedules and mobility
information help locate related activities \cite{fortes2018,bejarano2021},
while disruptions may redirect them \cite{zhang2021}. Propagation varies
with morphology, terrain, transmitter geometry and blockage
\cite{levie2021}. Advance preparation for Cell-on-Wheels (COW) deployment therefore requires a
spatial representation that connects activity descriptions with
radio-support assessment. Fig.~\ref{fig:context-to-cow} illustrates this context-to-area workflow.

Existing studies associate event information with network elements
\cite{fortes2018} and combine event context with cell traffic for
forecasting \cite{bejarano2021}. RadioMapSeer provides transmitter-specific
radio maps \cite{radiomap_dataset2024}; UrbanEMF provides city-scale and
local per-base-station received-signal-strength maps \cite{urbanemf2026}.
Sionna Large Radio Maps supports large-scale propagation simulation
\cite{sionnaLrm2025}. Service-area screening must map activity semantics to spatial demand and
combine that demand with relative propagation support over the same areas. Screening must preserve activity locations and local propagation variation.

We present EdgeNet City, an online framework that grounds free-form
natural-language inputs in geographic activity representations and
path-gain data for COW assessment. The framework separates
event-dependent demand construction from event-independent propagation
computation. Its two contributions are as follows.

\Needspace{4\baselineskip}
\emph{1) Event-to-area screening algorithm:} We introduce a
catalog-referenced activity representation that separates semantic selection
from deterministic spatial computation. A constrained large language model
(LLM) selects activity roles, geographic features and fixed demand profiles;
validation enforces admissible combinations. The selected references
determine prescribed demand on H3 service areas \cite{h3docs2026}, where the
algorithm combines demand with relative propagation support and record
availability. The expanded representation improves spatial agreement in the matched
20-event panel. With need-aware ranking and a three-cell diagnostic budget
fixed, it increases mean Reference-need Capture by 25.44 percentage points
over Venue-only.

\emph{2) Reusable propagation resource:} We construct and publicly release
the Greater London path-gain dataset \cite{greater_london_path_gain_2026}.
Transmitter- and radio-profile-specific gains are retained with receiver
geometry and provenance for local retrieval and reaggregation. A Tile6 case
demonstrates reuse across geographic windows, H3 resolutions and transmitter
subsets without new ray tracing (RT).

%% file: sections/02_system_model_and_problem_formulation.tex
\section{System Model and Problem Formulation}
\label{sec:system-model}

\subsection{Event Information and Activity Representation}
\label{sec:event-representation}

The inputs are a validated event $e$, source text $O_e$, a geographic
catalog $\mathcal M_e$, an allowed-reference registry $\mathcal C_e$ and a
precomputed path-gain resource. Geographic and temporal context supplies the
trusted event footprint $F_e$, the coordinate used for local propagation
preparation and the impact window $\tau_e$. The catalog contains the event
anchor and surrounding features; the registry lists permitted references,
including demand profiles.

An activity zone $z=(r_z,f_z,p_z)$ links an activity role $r_z$ to a
catalog-feature reference $f_z$ and a demand-profile reference $p_z$. The
feature reference supplies geometry $F_z$; the profile reference supplies
fixed demand parameters. Valid, compatible combinations form $\mathcal A_e$.
Event interpretation selects a finite, possibly empty activity
representation:

\begin{equation}
\mathcal Z_e=\{z_1,\ldots,z_{K_e}\}\subseteq\mathcal A_e,
\qquad K_e=|\mathcal Z_e|.
\label{eq:activity-representation}
\end{equation}

Each feature--profile pair is selected at most once. The supplied planning
limit bounds the number of distinct non-anchor feature references.

\subsection{Demand and Propagation on Common Service Areas}
\label{sec:shared-assessment-domain}

The event footprint and selected activity geometries define the H3 region
$\mathcal H_e$, whose cells are the service areas. A zone can contribute to
multiple cells, and a cell can receive demand from multiple zones. For fixed
activity geometries and impact window $\tau_e$, the dimensionless
event-demand increments for low, base and high scenarios,
$\omega\in\Omega=\{\mathrm L,\mathrm B,\mathrm H\}$, satisfy

\begin{equation}
0\leq d_{h,\tau_e}^{\mathrm L}
\leq d_{h,\tau_e}^{\mathrm B}
\leq d_{h,\tau_e}^{\mathrm H}\leq1,
\qquad h\in\mathcal H_e.
\label{eq:demand-field-domain}
\end{equation}

Zero denotes no assigned event-demand increment.

The path-gain resource $\mathcal D_{\mathrm{pg}}$ supplies path-gain data for each branch, indexed by transmitter cohort $o$ and radio
profile $q$. All branches share the demand fields. When the local
propagation scale is valid, cells satisfying the record-availability and
usability requirements form the eligible domain:

\begin{equation}
\begin{aligned}
\mathcal H^{\mathrm{eligible}}_{e,o,q}&\subseteq\mathcal H_e,\\
s_{h,o,q}&\in[0,1],
\quad h\in\mathcal H^{\mathrm{eligible}}_{e,o,q}.
\end{aligned}
\label{eq:propagation-field-domain}
\end{equation}

Larger $s_{h,o,q}$ denotes stronger relative propagation support on the fixed
local, profile-specific scale. All demand scenarios within a branch share
its support field and eligible domain. The event index is omitted where
unambiguous.

\subsection{Screening Task and Required Output}
\label{sec:screening-formulation}

Screening prioritizes service areas where stronger prescribed demand
coincides with weaker relative propagation support. A branch is available
when planning requirements are met, the assessment grid, propagation inputs
and normalization scale are valid, and the eligible domain is non-empty. The
output comprises a screening state $\widehat y_{e,o,q}$ and an optional
service-area cell $\widehat h_{e,o,q}$:

\begin{equation}
\begin{aligned}
\widehat y_{e,o,q}&\in\mathcal Y,\\
\widehat h_{e,o,q}&\in
\begin{cases}
\mathcal H^{\mathrm{eligible}}_{e,o,q}\cup\{\varnothing\},
&\text{available branch},\\
\{\varnothing\},&\text{otherwise},
\end{cases}
\end{aligned}
\label{eq:screening-output-domain}
\end{equation}

The state set $\mathcal Y$ contains \textsc{Assessment Unavailable},
\textsc{Priority Flagged}, \textsc{Further Assessment} and \textsc{No
Priority Flagged}. An unavailable branch returns \textsc{Assessment
Unavailable} with no cell. For available branches, \textsc{Priority Flagged}
and \textsc{Further Assessment} return an eligible cell for priority
assessment and monitoring, respectively. \textsc{No Priority Flagged}
returns no cell and applies within the eligible domain.

%% file: sections/03_methodology.tex
\section{Event-to-Area Screening Algorithm}
\label{sec:methodology}

EdgeNet City combines event-dependent activity demand and reusable propagation
support on the same H3 service-area grid (Fig.~\ref{fig:method-dataflow}).

\subsection{From Event Context to Spatial Demand Scenarios}
\label{sec:event-trigger}
\label{sec:investigator}

The upstream Trigger forms event $e$ through one structured language-model
call followed by deterministic source-reference and location checks.
The Planner input $X_e$ combines the event description and source text, trusted
geographic and temporal context, catalog attributes and allowed references.
The Planner selects role--feature--profile combinations for activities
relevant to $X_e$ during $\tau_e$ in at most one call:

\begin{equation}
\begin{aligned}
\Pi_e&=\mathcal L_\theta(X_e),\\
\widetilde\Pi_e&=\mathcal V(\Pi_e;\mathcal M_e,\mathcal C_e).
\end{aligned}
\label{eq:event-semantic-mapping}
\end{equation}

Here $\mathcal L_\theta$ denotes structured language-model inference with
fixed pretrained parameters $\theta$. Validator $\mathcal V$ enforces the
reference, compatibility and uniqueness conditions in
Section~\ref{sec:event-representation} and a limit of 12 distinct non-anchor
features. A rejected plan, $\widetilde\Pi_e=\bot$, returns
\textsc{Assessment Unavailable} with $\widehat h_{e,o,q}=\varnothing$. An accepted plan yields $\mathcal
Z_e=\operatorname{Zones}(\widetilde\Pi_e)$.
A saved Gtech Community Stadium plan assigns Kew Bridge the station-exit
role and station-arrival profile; its geometry locates the demand kernel,
and the profile fixes its radius and scenario amplitudes.

\label{sec:spatial-demand}
The grid $\mathcal H_e$ contains resolution-9 H3 cells whose centers lie
within the local-metric bounding box of $F_e$ and the selected $F_z$,
expanded by 750\,m. The domain must contain 7--120 cells and lie within the
supplied map extent; otherwise the algorithm returns \textsc{Assessment
Unavailable} before propagation scoring.

Let $A_h$ and $c_h$ denote the cell polygon and centroid, and let
$\delta(c_h,F_z)$ be the shortest local-metric distance to the zone
geometry. For radius $R_z>0$ and amplitude $m_z^\omega\ge0$, the influence
kernel and demand increment are

\begin{equation}
\begin{split}
\kappa_z(h)&=\begin{cases}
\min\{1,|A_h\cap F_z|/|A_h|\},&\text{polygonal venue},\\
[1-\delta(c_h,F_z)/R_z]_+,&\text{otherwise},
\end{cases}\\
d_{h,\tau_e}^{\omega}&=\operatorname{clip}_{[0,1]}
\left(\sum_{z\in\mathcal Z_e}m_z^{\omega}\kappa_z(h)\right).
\end{split}
\label{eq:normalised-demand}
\end{equation}

Vertical bars denote local-metric polygon area or set cardinality;
$[x]_+=\max\{0,x\}$ and
$\operatorname{clip}_{[0,1]}(x)=\min\{1,\max\{0,x\}\}$. The demand-profile
catalog fixes $R_z$ and the ordered amplitudes $0\le m_z^{\mathrm L}\le
m_z^{\mathrm B}\le m_z^{\mathrm H}$. For the evaluated scenarios, the
configured $(R_z;m_z^{\mathrm L},m_z^{\mathrm B},m_z^{\mathrm H})$ values,
with radius in meters, are venue $(350;0.45,0.65,0.85)$, station exit
$(300;0.30,0.50,0.70)$, event entrance $(200;0.35,0.55,0.75)$ and walking
corridor $(200;0.25,0.45,0.65)$. The 350\,m radius applies to the
venue-point fallback; polygonal venues use area overlap. Non-negative
kernels, ordered amplitudes and monotone clipping establish
\eqref{eq:demand-field-domain}. An accepted empty activity set gives a zero
demand field.

\subsection{Reusable City-Scale Propagation Resource}
\label{sec:path-gain-dataset}

The released Greater London path-gain dataset
\cite{greater_london_path_gain_2026} stores event-independent linear gains
by transmitter and radio profile. It retains receiver geometry, provenance
and manifests for local retrieval, record checks and reaggregation.
Its 449,437,201,731 gain records link
22,678 transmitter proxies to 171,549,960 receiver faces across
four radio access technologies (RATs): Global System for Mobile
Communications (GSM), Universal Mobile Telecommunications System (UMTS),
Long-Term Evolution (LTE) and New Radio (NR).

Within the December 2025 Greater London boundary \cite{onsLondon2025},
crowdsourced cellular observations \cite{opencellid2026} are grouped by
network code and RAT and clustered within 50\,m. A sample-count-weighted
observed medoid represents each cluster and defines its transmitter proxy. Sionna Large Radio Maps \cite{sionnaLrm2025} constructs
adaptive tiled scenes from OpenStreetMap geometry and terrain
\cite{osmCopyright,eaDtm2022,mapzenTerrain}. Sionna RT \cite{sionnart2025}
computes face-area-averaged linear gains, accounting for line of sight,
specular reflection and slab transmission. Transmitters and receivers are
placed 15\,m above terrain and 1.5\,m above ground, respectively. The
simulations use $1\times1$ vertically polarized isotropic antennas and a
maximum interaction depth of 10, with at least $4.0\times10^7$ rays per
transmitter and a nominal $1.0\times10^9$ rays per tile.
Fig.~\ref{fig:dataset-overview} shows the city-scale extent.
\input{figures/fig_dataset_overview}
\input{figures/float_propagation_triptych}

\subsection{Event-Local Propagation Support}
\label{sec:path-gain}

A prepared event bundle $b(e)$ supplies records within its declared H3
domain. Its receiver tiles, selected transmitters and normalization anchors
are fixed before evaluation and retained across queries. For each branch,
bundle preparation selects the three nearest transmitters with computed gain
records for all required receiver tiles. Great-circle distance is measured
from the supplied event coordinate, with ties resolved by ascending
lexicographic identifier. A branch is unavailable when fewer than three
candidates exist.

Let $g_{u,j}\ge0$ denote the ray-traced linear gain from transmitter instance
$u$ to receiver face $C_j$, whose surface area is $A_j>0$. The index $j$
identifies a face by its tile and local index. Each whole face is assigned
to the H3 cell containing its centroid. The expected set $\mathcal
J^{\mathrm{exp}}_{u,h}$ contains faces assigned to $h$ for which records
from $u$ are expected. Receiver geometry determines this set and the face
areas before gain lookup. The availability indicator $I_{u,j}$ is one for a
finite solver value, including a computed zero, and zero for a missing
value. For a non-empty expected set, area completeness and available-face
mean gain are

\begin{equation}
\begin{aligned}
\gamma_{u,h}^{o,q}&=
\frac{\sum_{j\in\mathcal J_{u,h}^{\mathrm{exp}}}A_jI_{u,j}}
{\sum_{j\in\mathcal J_{u,h}^{\mathrm{exp}}}A_j},\\
\bar g_{u,h}^{o,q}&=
\frac{\sum_{j\in\mathcal J_{u,h}^{\mathrm{exp}}:I_{u,j}=1}A_jg_{u,j}}
{\sum_{j\in\mathcal J_{u,h}^{\mathrm{exp}}}A_jI_{u,j}}.
\end{aligned}
\label{eq:h3-propagation-features}
\end{equation}

The mean is undefined when no expected face has an available value. The
usable transmitter set $\mathcal U^{\mathrm{use}}_{h,o,q}$ contains
transmitters with defined means and $\gamma\ge0.95$. If none qualify, it
contains those with defined means and $0.80\le\gamma<0.95$. A cell is
eligible when all manifest-required transmitter--H3 summaries are present
and $\mathcal U^{\mathrm{use}}_{h,o,q}$ is non-empty. An existing summary
with an undefined mean counts as present; a physically absent required
summary excludes the cell.

Under fixed transmit power and receiver assumptions, larger desired-link
path gain corresponds to stronger received signal power. For each declared H3 cell and branch, bundle preparation retains the
strongest positive, defined mean gain among selected transmitters, when one
exists. These retained gains are converted to dB and pooled across cohorts before
event-local eligibility is applied. At least 20 pooled entries are required
to compute the fifth- and 95th-percentile anchors $G^5_{b,q}$ and
$G^{95}_{b,q}$ by linear interpolation. Finite, strictly ordered anchors
define a valid scale; otherwise the branch is unavailable. Normalized gain
and eligible-cell support are

\begin{equation}
\begin{aligned}
\phi_{b,q}(g)&=\begin{cases}
0,&g=0,\\
\operatorname{clip}_{[0,1]}\!\left(
\dfrac{10\log_{10}g-G^5_{b,q}}{G^{95}_{b,q}-G^5_{b,q}}\right),&g>0,
\end{cases}\\
s_{h,o,q}&=\max_{u\in\mathcal U^{\mathrm{use}}_{h,o,q}}
\phi_{b(e),q}(\bar g_{u,h}^{o,q}).
\end{aligned}
\label{eq:p0-support-mapping}
\end{equation}

The transmitter attaining maximum support is retained as the winner. Ties
favor higher completeness, then the lexicographically smaller identifier.

\subsection{Demand--Support Screening}
\label{sec:cow-recommendation}

For an available branch and $h\in\mathcal H^{\mathrm{eligible}}_{e,o,q}$,
the priority index and its base/high maxima are

\begin{equation}
\begin{aligned}
n_{h,o,q}^{\omega}&=d_{h,\tau_e}^{\omega}(1-s_{h,o,q}),\quad \omega\in\Omega,\\
M_{e,o,q}^{\omega}&=\max_{h\in\mathcal H^{\mathrm{eligible}}_{e,o,q}}
n_{h,o,q}^{\omega},\quad \omega\in\{\mathrm B,\mathrm H\}.
\end{aligned}
\label{eq:cow-deployment-need}
\end{equation}

With the other quantity fixed, the index is non-decreasing in demand and
non-increasing in support. Shared support and eligibility, together with
\eqref{eq:demand-field-domain}, give $M^B_{e,o,q}\le M^H_{e,o,q}$. Eligible
cells are ranked by descending $n^\omega$, descending winning-transmitter
completeness and ascending H3 identifier. Let $h^\omega_{e,o,q}$ denote the
first-ranked cell for $\omega\in\{\mathrm B,\mathrm H\}$. Base-scenario
diagnostic rankings use the same ordering.

For radio profile $q$ and dimensionless threshold $\tau_{\mathrm{cow},q}$,
the screening state is

\begin{equation}
\renewcommand{\arraystretch}{1.2}
\begin{aligned}
&\widehat y_{e,o,q}=\\
&\left\{\begin{array}{@{}l@{\enspace}l@{}}
\textsc{Assessment Unavailable},&\text{branch unavailable},\\
\textsc{Priority Flagged},&M^{\mathrm B}_{e,o,q}\geq\tau_{\mathrm{cow},q},\\
\textsc{Further Assessment},&M^{\mathrm B}_{e,o,q}<\tau_{\mathrm{cow},q}\leq M^{\mathrm H}_{e,o,q},\\
\textsc{No Priority Flagged},&M^{\mathrm H}_{e,o,q}<\tau_{\mathrm{cow},q}.
\end{array}\right.
\end{aligned}
\label{eq:deployment-recommendation-state}
\end{equation}

The last three cases apply to available branches. \textsc{Priority Flagged}
returns $\widehat h_{e,o,q}=h^B_{e,o,q}$; \textsc{Further Assessment}
returns $\widehat h_{e,o,q}=h^H_{e,o,q}$. The other states return no cell.
Unavailable branches have no computed maxima or rankings.

The online prototype combines event-dependent demand construction with
offline propagation summaries within prepared domains, allowing queries
without repeating RT. Fig.~\ref{fig:h3-priority-map} presents prototype
screen captures under uniform demand, showing how cohort-specific support
produces different screening states on the same assessment grid. Under
uniform demand, the maxima are 0.4819 for cohort A and 0.4066 for cohort B
at threshold 0.45.

%% file: figures/fig_dataset_overview.tex
\begin{figure}[!t]
    \centering
    \includegraphics[width=\columnwidth,height=50.5mm,keepaspectratio]{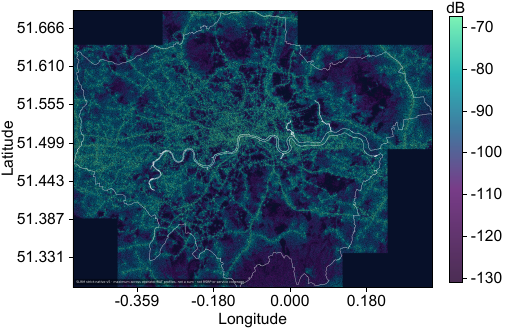}
    \caption{City-scale overview of the Greater London path-gain dataset. At each receiver face, the visualization takes the maximum linear gain
across available cohort--RAT records spanning GSM, UMTS, LTE, and NR, then
converts it to dB. The display uses the 2nd and 98th percentiles as scale limits;
per-transmitter records remain available in the archive.}
    \label{fig:dataset-overview}
\end{figure}

%% file: figures/float_propagation_triptych.tex
\begin{figure*}[!t]
\centering
\includegraphics[width=0.93\textwidth]{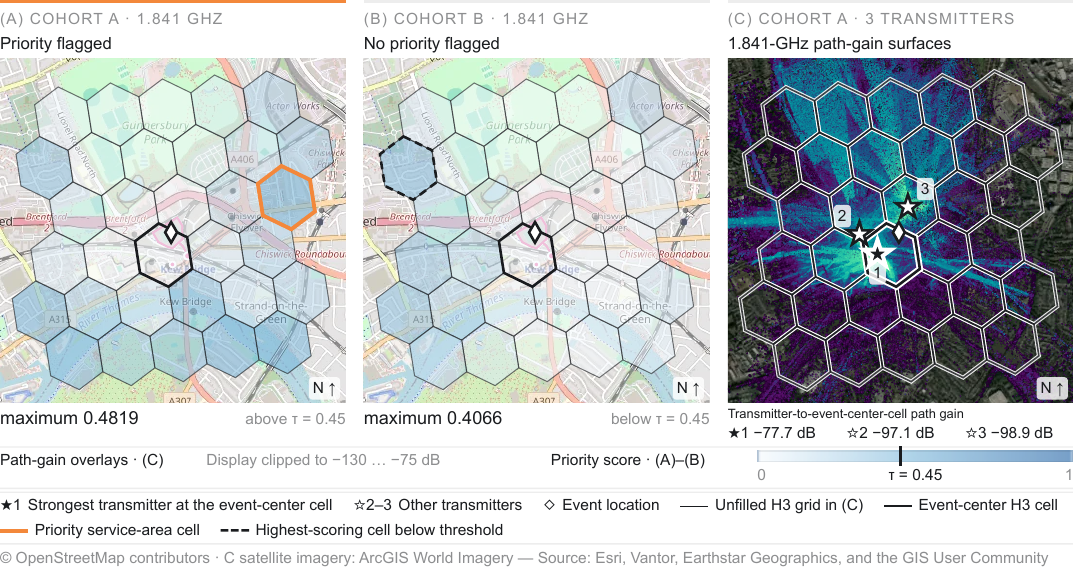}
\caption{Online prototype results at Gtech Community Stadium under uniform demand
$d=1$. Panels (A) and (B) use different transmitter cohorts with the same event,
32-cell resolution-9 H3 grid, 1.841-GHz profile and screening threshold of
0.45. Orange marks
the returned cell in (A); the dashed below-threshold maximizer in (B) is
shown for visualization and is not returned. Panel (C) overlays sampled gains from existing transmitter proxies on
satellite imagery, drawing stronger samples last. The three reported gains
refer to the event-center cell.}
\label{fig:h3-priority-map}
\end{figure*}

%% file: sections/04_results.tex
\section{Evaluation and Discussion}
\label{sec:results-evaluation}
\input{figures/tab_activity_decomposition}

\subsection{Evaluation Setup}

The matched evaluation compares Venue-only, Rule GIS (geographic information
system), Ordinary LLM and Constrained Planner on 20 scheduled events across
six London venue--radio environments used during development. All methods
share event information, a compatible role--feature--profile catalog,
validation and the deterministic screening core.

Venue-only serves as the single-feature baseline and selects one venue
feature. For these pre-event cases, Rule GIS selects at most one compatible
feature per activity role, excluding closed or conflicting features. It
orders candidates by distance bucket relative to the event geometry---within
250\,m, then 250--500\,m---and then by lexical reference. Both methods use
the trusted anchor when the venue catalog is empty. Ordinary LLM uses a
plain structured-planning prompt; Constrained Planner uses the production
Planner prompt. Both request \texttt{gpt-6.1-sol}/medium.

Each plan is evaluated for two LTE cohorts, producing 160 outputs across
four methods; events remain the planning units. Cohorts A/B correspond to
network codes 15/30 at 1.841\,GHz. The screening-index threshold is 0.45,
the event-impact window is 60\,min, and resolution-9 H3 cells have a global
mean area of $0.105\,\mathrm{km}^2$ \cite{h3docs2026}.

All methods use the same technical-review reference, adjudicated from
candidate inventories without independent, output-blind spatial
re-annotation from source materials. The full reference contains 54 zones;
the reachable subset contains 39 exact catalog-expressible tuples. The same
predictions are evaluated against both reference scopes.

\subsection{Spatial Agreement of Activity Representations}

This comparison tests whether expanded activity representations improve the
spatial alignment of prescribed demand. The saved reference-evaluation grid
contains resolution-9 H3 cells selected by center containment in a bounding
box enclosing the event footprint, historical prediction geometries and the
full reference, with 750\,m padding. Predicted and reference zones use the
same prescribed role profiles. Their uncapped per-zone base-demand
contributions are summed and normalized to unit mass on this common grid.
\mbox{Spatial overlap is $\sum_h\min(p_h^{\mathrm{pred}},p_h^{\mathrm{ref}})$.}
\mbox{Role-multiset F1 measures} role counts, retaining repeated roles
independently of place identity.

Relative to Venue-only, expanded representations increase mean overlap
from 0.4929 to 0.7203 against the full reference and from 0.6226 to 0.8308
against the reachable reference (Table~\ref{tab:activity-decomposition}).
Full-reference overlap improves in all 20 paired events; mean gains across
environments range from 0.0369 to 0.3911. Reachable-reference overlap
improves in 16 events and decreases in four, all at Emirates. Rule GIS and
both LLM planners select identical role--feature--profile multisets in all
20 cases. Their shared gains reflect the expanded activity representation
and prescribed spatial mapping.

\subsection{Fixed-Budget Service-Area Prioritization}

A $2\times2$ diagnostic separates representation effects from ranking
effects by crossing Venue-only and Rule GIS demand fields with demand-only
and need-aware ranking. Each condition selects three eligible cells from a
common selector grid. The grid uses resolution-9 center containment in a
bounding box enclosing the event footprint and all catalog geometries, with
750\,m padding. Demand-only ranking orders cells by descending $d_h^{\mathrm
B}$; need-aware ranking uses $n_h^{\mathrm B}=d_h^{\mathrm B}(1-s_h)$. Both
use the tie-breaking rule in Section~\ref{sec:cow-recommendation}.

Demand Capture sums the selected reference mass using its original
normalization. Reference-need Capture sums the selected
$p_h^{\mathrm{ref}}(1-s_h)$ after normalization within the selector grid.
This three-cell budget defines the diagnostic ranking; the screening
algorithm returns a state and an optional single service area. Under the
prescribed demand scenarios, all 160 outputs are \textsc{No Priority
Flagged}. Every domain cell is eligible, and the maximum high-scenario
priority index is 0.234062, below the fixed threshold of 0.45.
\input{figures/tab_fixed_budget_components}

The expanded representation improves both mean Capture measures under both
ranking rules (Table~\ref{tab:fixed-budget-components}). With need-aware
ranking fixed, it increases mean Reference-need Capture by 25.44 percentage
points and mean Demand Capture by 3.26 percentage points over Venue-only.

\Needspace{6\baselineskip}
With the expanded representation fixed, need-aware ranking increases mean
Reference-need Capture by 41.46 \mbox{percentage points} and decreases mean Demand
Capture by 53.30 \mbox{percentage points} relative to demand-only ranking. It prioritizes demand with weaker propagation support.

\subsection{Role of Spatial Propagation}
\label{sec:spatial-propagation-control}

A controlled replay isolates how cell-level propagation support affects
screening states. For the same 20 events and two LTE cohorts, it sets $d=1$
and holds transmitters, grid, anchors, eligibility and threshold fixed.
Cell-level support yields 30 \textsc{Priority Flagged} outputs across 40
event--cohort branches. Replacing it with its unweighted mean over eligible
cells yields four. Spatial averaging removes threshold exceedances
in 26 of the 40 event--cohort branches.

\balance
\subsection{Reuse of Path-Gain Data}

The Tile6 case tests whether saved face-level gains and receiver geometry
support reuse across spatial windows, H3 resolutions and transmitter subsets
without new RT. The case uses two transmitter proxies, one LTE profile and
six 1\,km square windows selected from geometry alone. Two windows have the
lowest and highest building-projection fractions on a fixed $3\times3$
interior grid; the other four are centered at the midpoints of the scene's
bounding-box edges.

Original face centroids determine window membership and H3 assignment, with
whole-face areas retained. Gains are reaggregated at H3 resolutions 8 and 9.
Completeness uses each full cell's expected-area denominator; mean gain uses
the available-face area within the window.

Face identifiers match between the gain records and receiver geometry.
Recovered coordinates and recomputed areas agree with the saved geometry. For every window and proxy,
cell means match independent face-by-face calculations, and gain--area
totals are conserved across resolutions within numerical tolerance. At resolution 9, the same records yield per-cell maximum mean gains for
either proxy alone and for the pair.

%% file: figures/tab_activity_decomposition.tex
\begin{table}[!t]
\centering
\footnotesize
\setlength{\tabcolsep}{2.6pt}
\caption{Matched 20-event comparison against the reused technical-review reference. The full and reachable references contain 54 and 39 role instances, respectively. $\mu F_1$ and $m F_1$ denote role-multiset micro and case-macro F1,
respectively; $O$ denotes mean unit-mass spatial overlap.}
\label{tab:activity-decomposition}
\begin{tabular}{@{}lrrrrrr@{}}
\hline
& \multicolumn{3}{c}{Full} & \multicolumn{3}{c}{Reachable} \\
Method & $\mu F_1$ & $m F_1$ & O & $\mu F_1$ & $m F_1$ & O \\
\hline
Venue-only & 0.5405 & 0.5600 & 0.4929 & 0.6780 & 0.7083 & 0.6226 \\
Rule GIS & 0.7752 & 0.7614 & 0.7203 & 0.6842 & 0.6752 & 0.8308 \\
Ordinary LLM & 0.7752 & 0.7614 & 0.7203 & 0.6842 & 0.6752 & 0.8308 \\
Constrained Planner & 0.7752 & 0.7614 & 0.7203 & 0.6842 & 0.6752 & 0.8308 \\
\hline
\end{tabular}
\end{table}

%% file: figures/tab_fixed_budget_components.tex
\begin{table}[!t]
\centering
\footnotesize
\setlength{\tabcolsep}{3pt}
\renewcommand{\arraystretch}{1.08}
\caption{Mean paired differences in Capture (percentage points), with three selected
cells per condition.}
\label{tab:fixed-budget-components}
\begin{tabular}{@{}p{0.53\columnwidth}rr@{}}
\hline
Contrast (fixed factor) &
\shortstack{Demand\\Capture\\($n=40$)} &
\shortstack{Reference-need\\Capture\\($n=28$)} \\
\hline
Expanded $-$ Venue-only\newline (demand-only ranking) &
$+19.22$ & $+20.68$ \\
Expanded $-$ Venue-only\newline (need-aware ranking) &
$+3.26$ & $\mathbf{+25.44}$ \\
Need-aware $-$ demand-only\newline (Venue-only representation) &
$-37.34$ & $+36.70$ \\
Need-aware $-$ demand-only\newline (expanded representation) &
$-53.30$ & $\mathbf{+41.46}$ \\
\hline
\end{tabular}
\par\vspace{2pt}
\begin{minipage}{\columnwidth}
\footnotesize
Expanded denotes Rule GIS; both LLM methods match its activity selections
and Capture values. The 20 events and two cohorts provide 40 paired event--cohort comparisons.
Reference-need Capture is not applicable (NA) for the 12 pairs with zero
total reference-need mass.
The selector grid retains a mean of 0.9591 of the original reference
mass (range 0.7955--1.0000). Demand Capture retains the original
normalization; Reference-need Capture is normalized within the selector grid.
\end{minipage}
\end{table}

%% file: sections/06_conclusion.tex
\par\addvspace{8pt}
\noindent\begin{minipage}[b]{\columnwidth}
\setlength{\parindent}{1em}
\section{Conclusion}

  In the matched 20-event panel, the expanded representation improves spatial
agreement. With need-aware ranking and the three-cell diagnostic budget
fixed, it increases mean Reference-need Capture by 25.44 percentage points
over Venue-only. EdgeNet City connects natural-language context to spatial demand and
propagation-based prioritization for COW assessment. Its online prototype
and released Greater London dataset make this workflow executable.
\end{minipage}

%% file: sections/08_references.tex
\begingroup
\emergencystretch=1.5em

\endgroup